\documentclass[aps,prl,twocolumn,showpacs,superscriptaddress,nofootinbib,nobibnotes]{revtex4-1}

\usepackage{breqn}
\usepackage{graphicx}
\usepackage{dcolumn}   
\usepackage{bm}        
\usepackage{amssymb}   

\usepackage[breaklinks=true,colorlinks=true,linkcolor=blue,urlcolor=blue,citecolor=blue]{hyperref}

\usepackage{float}

\begin{document}

\title{Turbulence suppression by energetic particle effects in modern optimized stellarators}

\author{A.~Di Siena} 
\affiliation{The University of Texas at Austin 201 E 24th St 78712 Austin Texas USA}
\affiliation{Max Planck Institute for Plasma Physics Boltzmannstr 2 85748 Garching Germany} 
\author{A.~Ba\~n\'on~Navarro}
\affiliation{Max Planck Institute for Plasma Physics Boltzmannstr 2 85748 Garching Germany}
\author{F.~Jenko} 
\affiliation{Max Planck Institute for Plasma Physics Boltzmannstr 2 85748 Garching Germany}

\begin{abstract}

Turbulent transport is known to limit the plasma confinement of present-day optimized stellarators. To address this issue, a novel method to strongly suppress turbulence in such devices is proposed, namely the resonant wave-particle interaction of supra-thermal particles - e.g., from ion-cyclotron-resonance-frequency (ICRF) heating - with turbulence-driving microinstabilities like ion-temperature-gradient (ITG) modes. The effectiveness of this mechanism is demonstrated via analytic theory and large-scale gyrokinetic simulations, revealing an overall turbulence reduction by up to $80\%$. These results hold the promise of new and still unexplored stellarator scenarios with enhanced confinement and improved performance, essential for achieving burning plasmas in future devices.

\end{abstract}

\pacs{52.65.y,52.35.Mw,52.35.Ra}

\maketitle


{\em Introduction.} Turbulent transport, generated by pressure-gradient-driven microinstabilities and inducing significant energy and particle losses, is often a key limiting factor for the performance of magnetic confinement fusion devices. Historically, most investigations on plasma turbulence were carried out for tokamak experiments, while stellarators were usually limited by neoclassical (collisional) transport.  However, a large number of experimental \cite{Stroth_1998,Motojima_2003,Deng_2015} and theoretical studies \cite{Xanthopoulos_2007,Xanthopoulos_2014} have demonstrated the critical role played by turbulence in stellarator devices when neoclassical transport is reduced. In this context, we note, in particular, recent observations at Wendelstein 7-X (W7-X) \cite{Nuhrenberg_1986,Klinger_2013}, 
which show that ion-temperature-gradient (ITG) \cite{Romanelli_1989}
driven turbulent transport clearly exceeds its neoclassical counterpart
\cite{Wolf_2017,Klinger_2019}. These results emphasize the great importance of identifying mechanisms able to systematically suppress turbulent transport in modern optimized stellarators.

In the present Letter, we propose a novel method to strongly suppress turbulence in such devices, namely the resonant wave-particle interaction of supra-thermal (fast) ions with turbulence-driving microinstabilities   \cite{DiSiena_NF2018,DiSiena_PoP2019}. The beneficial role of this mechanism is demonstrated via nonlinear gyrokinetic simulations \cite{Brizard_2007} of W7-X. We show that fast ions - under appropriate conditions - significantly impact turbulence-induced energy losses, potentially leading to a considerable reduction of turbulent transport. This mechanism might therefore lead to a substantial increase in the bulk ion temperature profile peaking, with a consequential improvement of plasma confinement. The results are supported by analytic theory, which reveals that such turbulence reduction is maximized when the generated fast particle population simultaneously fulfills the following conditions: (i) steep temperature and rather flat density profiles, (ii) moderate temperatures (one order of magnitude larger than the bulk). Such conditions can be achieved by ion-cyclotron-resonance-frequency (ICRF) heating schemes \cite{DiSiena_NF2018}. These findings are particularly attractive for steady-state stellarator devices, providing an external tool - such as supra-thermal particles - to continuously reduce turbulent transport.

{\em Analytic model.} The fast ion effect we propose to exploit in optimized stellarator devices can be well captured by a quasi-linear model derived from the gyrokinetic framework \cite{Brizard_2007}. Gyrokinetics is a rigorous limit of kinetic theory, employed to study turbulence in strongly magnetized, weakly collisional  plasmas. It is based, in particular, on the assumptions of small fluctuation amplitudes and low frequency dynamics, which allow to remove various small (and irrelevant) space-time scales and one of the velocity space dimensions from the problem. In this model, the contribution of the energetic particles to the exponential growth of the plasma microinstabilities (here called $\gamma_f$) can be measured by studying the energy exchanged between the fast ions and the most unstable mode \cite{Hatzky_PoP2002,Manas_2015,DiSiena_NF2018,Novikau_2019}. Negative (positive) values of $\gamma_f$ indicate that the energetic particle species is taking (giving) energy from (to) the plasma micro-instabilities with a consequent damping (growth) of the mode. In normalized units, $\gamma_f$ can be written approximately as
\begin{equation}
\gamma_{f} \propto \int n_f\frac{\frac{a}{L_{n,f}}+\frac{a}{L_{T,f}}\left(E_f-\frac{3}{2}\right)}{\omega_k+\omega_{d,f}}e^{-E_f}E_f^{3/2}dE_f. 
\label{eq:gamma_f}
\end{equation}
For a full derivation of this expression, see Supplemental Material \cite{Supplemental}. In this relation,  $E_f = \left(v_\shortparallel^2+\mu B_0\right)$ is the energetic particle energy with $v_\shortparallel$ the velocity component parallel to the background magnetic field $B_0$ and $\mu$ the magnetic moment, $a/L_{T,f}$ and $a/L_{n,f}$ are the fast ion normalized logarithmic temperature and density gradients with $a$ the minor radius of the device, $\omega_k$ the frequency of the most unstable mode (with positive values denoting a mode propagating in the ion-diamagnetic direction) and $\omega_{d,f} = k_y\mathcal{K}_yT_f E_f/q_f$ the fast ion bi-normal drift frequency. Here,  $T_f$, $n_f$ and $q_f$ are, respectively, the energetic particle temperature, density (both normalized to the electron ones) and charge, $k_y$ is the bi-normal wave vector of the selected mode and 
%
\begin{equation}
\mathcal{K}_y = - \frac{\left( \mathbf{B_0} \times \mathbf{\nabla} B_0\right) \cdot \hat{y}}{B_0^2},
\label{eq:binormal}
\end{equation}
is the bi-normal curvature term, where $\hat{y}$ denotes the unit vector along the bi-normal direction. 
Eq.~(\ref{eq:gamma_f}) reveals that a resonance ($\omega_k + \omega_{d,f} = 0$) occurs in the fast ion contribution to the linear growth rate whenever the frequency of the unstable mode $(\omega_k)$ matches the bi-normal drift frequency of the energetic particles $(\omega_{d,f})$. As a direct consequence of Eq.~(\ref{eq:gamma_f}), the term $\mathcal{K}_y$ must be negative for modes propagating in the ion diamagnetic direction - such as ITG modes - to fulfill the resonance condition. This constraint grants a central role to the bi-normal curvature term $\mathcal{K}_y$, which is exceptionally complex in stellarator geometries due to the sophisticated magnetic field configuration. 

Another observation from Eq.~(\ref{eq:gamma_f}) is that the term $a/L_{n,f}+a/L_{T,f}\left(E_f-3/2\right)$, commonly called the drive term, sets the sign of $\gamma_f$. It yields to negative values within $E_f < 3/2$ (or equivalently $v_\shortparallel^2+\mu B_0 < 3/2$), when $a/L_{T,f} > a/L_{n,f}$. Therefore, a negative (and hence beneficial) fast particle contribution to the linear ITG mode growth can be achieved \textit{only} when (i) $a/L_{T,f} > a/L_{n,f}$ and (ii) $\omega_k$ matches $\omega_{d,f}$ in the phase-space region where $E_f < 3/2$. These conditions are usually satisfied by energetic particles generated via ICRF heating \cite{DiSiena_NF2018}.

{\em Gyrokinetic simulations for optimized stellarators.} The impact of this resonant mechanism on the turbulent transport of stellarator devices is investigated through numerical simulations of W7-X in the high-mirror configuration, performed with the gyrokinetic code GENE \cite{Jenko_PoP2000,Goerler_JCP2011,Xanthopoulos_2014}. To reduce the massive computational cost of these simulations, the radially local approximation is used. We present results obtained (i) by flux-tube simulations, using the most unstable "bean-shaped" flux tube on the flux surface and (ii) by full-flux-surface simulations. The first approach assumes periodic bi-normal boundary conditions, thus enabling a Fourier decomposition of the perturbed quantities along this direction \cite{Dannert_2005}. This method allows us to restrict the numerical analyses only to the most relevant wave-numbers, significantly reducing the grid resolution required. However, the nontrivial magnetic field-line dependence of the turbulent transport in stellarator geometries can only be captured correctly by relaxing the periodic boundary assumption and treating the bi-normal direction in real space \cite{Xanthopoulos_2014}. This second approach is significantly more expensive and will be used to corroborate the flux-tube results.

The chosen simulation parameters, summarized in Tab.~\ref{table:parameters}, are inspired by realistic W7-X data \cite{Klinger_2019} and refer to experimental conditions in which the turbulence is driven by ITG modes.
\begin{table} 
\caption{\label{tab:parameters}Plasma parameters used for the flux-tube ("bean -shaped") and full surface simulations of a high-mirror W7-X configuration.}
\begin{ruledtabular}
\begin{tabular}{c c c c c c}
Species & $T$ & $n$&  $a/L_T$ & $a/L_n$\\
\hline
H & 1.0-30.0 & 0.04 & 18.0 & 0.0 \\
\hline
D & 1.0 & 0.96 & 2.5 & 0.0 \\
\end{tabular}
\end{ruledtabular}
\label{table:parameters}
\end{table}
The bulk plasma is composed by Deuterium (D) and electrons. The latter are mostly assumed to have an adiabatic response. Kinetic electron simulations are also performed throughout this Letter for the most unstable flux-tube setup to further validate the adiabatic electron results. Similar analyses are prohibitive currently for the full surface simulations due to the massive amount of computational resources required. A finite value for the logarithmic density gradient of $a/L_{n,e} = 0.5$ (same for each species), temperature gradient $a/L_{T,e} = 2.0$ and the ratio between ion-electron temperatures $T_i/T_e = 0.7$ are employed. The choice of this specific set of parameters is also motivated by the ones typically observed in the core region of W7-X. The supra-thermal ions are modelled with ICRF parameters of a minority heating scenario of Hydrogen (H) in D with $n_H/n_e = 6\%$, as recently proposed in Ref.~\cite{IAEA_2018} for the ICRF system in W7-X (currently under construction). Moreover, we consider $a/L_{Tf} = 18$, which is consistent with the realistic values usually observed in ICRF H-minority heating in tokamak experiments \cite{Bonanomi_2018,Di_Siena_2019}. Finally, the grid resolution in radial, bi-normal and parallel to the magnetic field line directions for
the flux-tube and full surface GENE simulations are respectively $(x,y,z) = (144,128,192)$ and $(x,y,z) = (250,32,128)$. A fixed resolution in the magnetic moment and parallel velocity $(\mu,v_\shortparallel) = (20,32)$ is used.

\begin{figure}
\begin{center}
\includegraphics[scale=0.48]{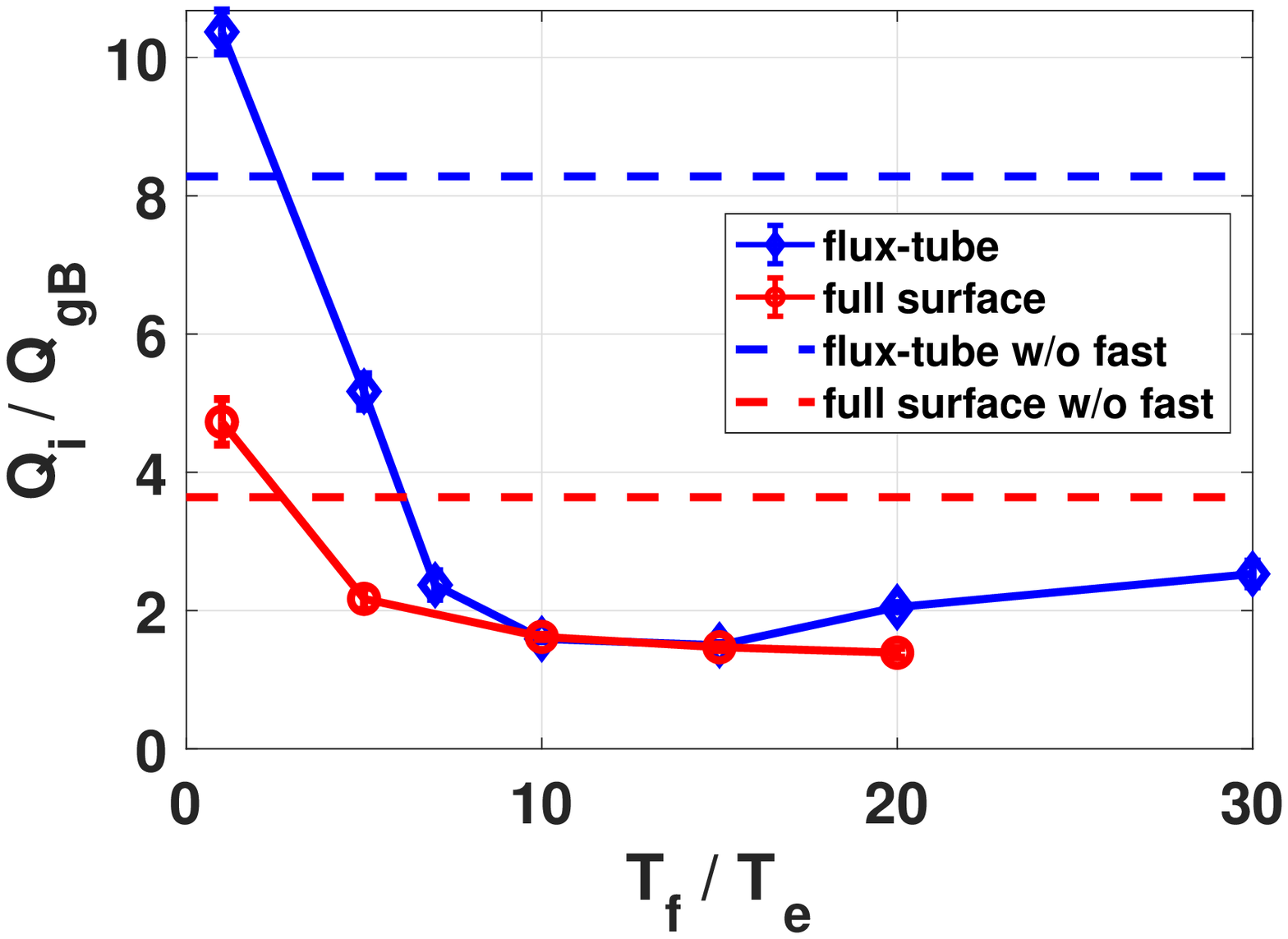}
\includegraphics[scale=0.48]{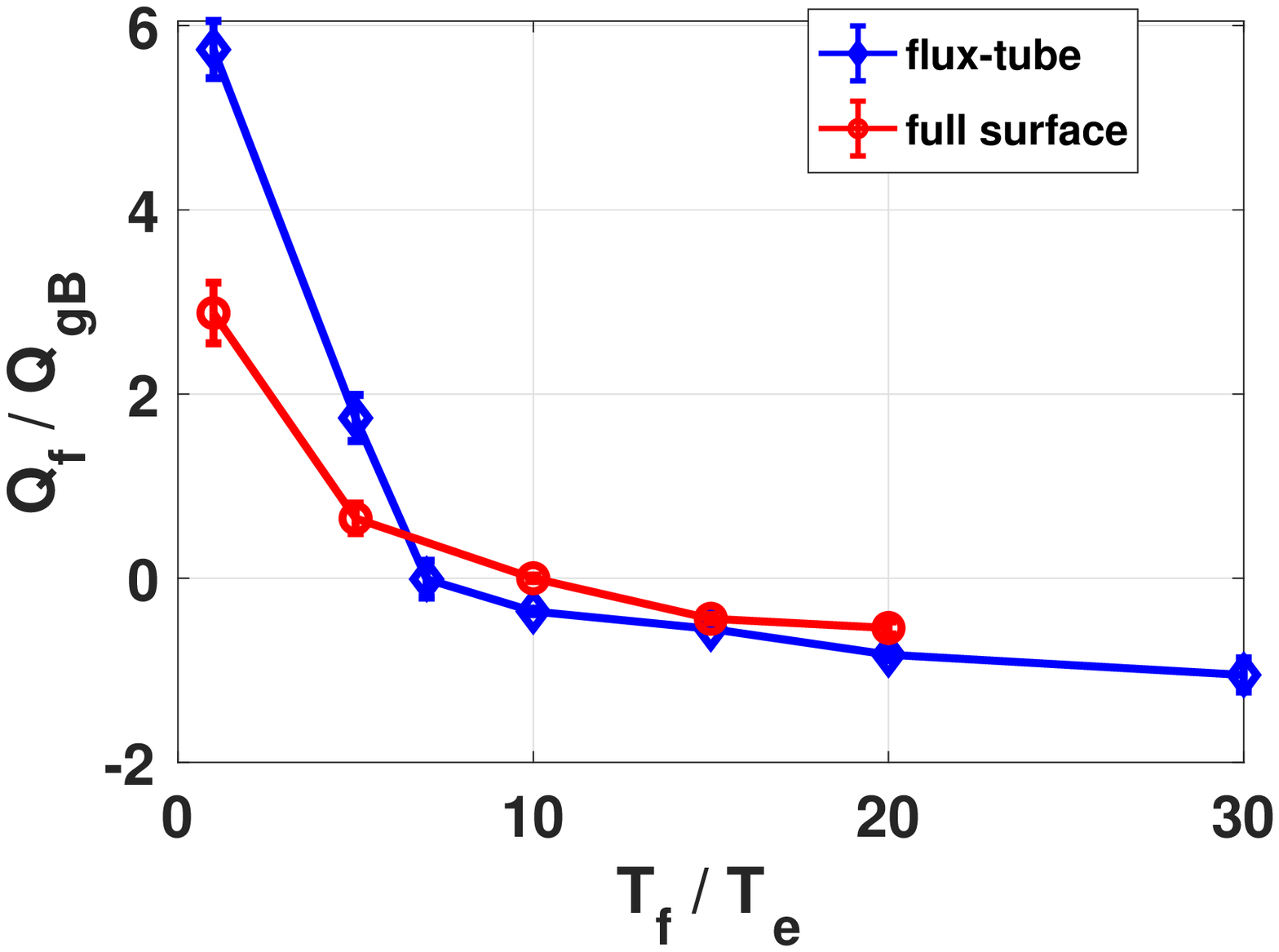}
\par\end{center}
\vspace{-0.7cm}
\caption{Nonlinear main ion (a) and energetic particle (b) heat fluxes in GyroBohm $(Q_{gB})$ units for different energetic particle temperatures $T_f/T_e$. The horizontal dotted lines denote the fluxes obtained without fast ions.}
\label{fig:fig1}
\end{figure}

{\em Turbulence reduction through fast particles.} As a first attempt to quantify the impact of the wave-particle resonance mechanism in W7-X (and similar optimized stellarator devices), the nonlinear energy fluxes are studied for different values of the energetic particle temperatures. In particular, the flux surface averaged thermal and supra-thermal ion heat fluxes - normalized to $Q_{gB} = n_i \, T_i \,c_s \,(\rho_s/a)^2$ - are shown in Fig.~\ref{fig:fig1}. Here, $\rho_s = (T_e / m_i)^{1/2} / \Omega_i$ represents the thermal gyroradius, with $\Omega_i = (q_i B_0) / (m_i c)$ the gyro-frequency, $c_s = \left(T_e/m_i\right)^{1/2}$ the sound speed, $m_i$ the main ion mass, $q_i$ the charge. 
A striking feature in Fig.~\ref{fig:fig1}a is the particularly strong dependence of the main ion fluxes on the externally heated H temperature. A pronounced reduction is found around $T_f \sim 10 T_e$ with $\sim 80\%$ turbulence suppression compared to the case with only thermal particles (dashed blue line in Fig~\ref{fig:fig1}a) for the most unstable flux-tube. Furthermore, in correspondence of such a meaningful ITG stabilisation, Fig.~\ref{fig:fig1}b reveals a change in the direction of the fast ion turbulent energy fluxes, which reverts from outward to inward. These results are not exclusive to the local setup, but extend to the more realistic full surface simulations, where a $\sim 55\%$ reduction (at $T_f / T_e = 10$) in the main ion energy flux is found again to be correlated to  inward energetic particle losses. Moreover, within the limit $T_f \sim T_e$, large supra-thermal particle fluxes are observed in Fig.~\ref{fig:fig1}a together with an increase in the anomalous main ion transport compared to the reference case without energetic particles.

\begin{figure}
\begin{center}
\includegraphics[scale=0.25]{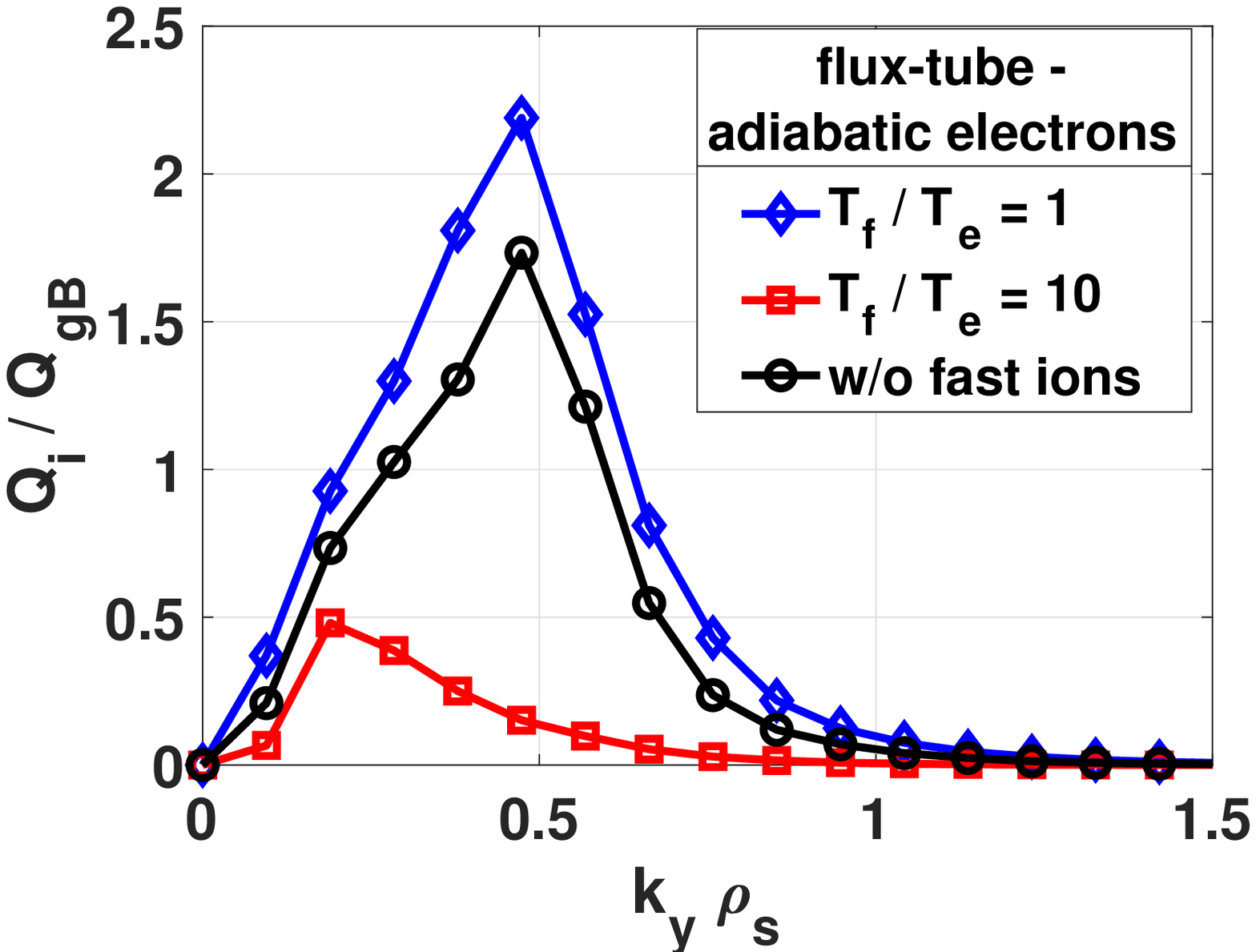}\includegraphics[scale=0.25]{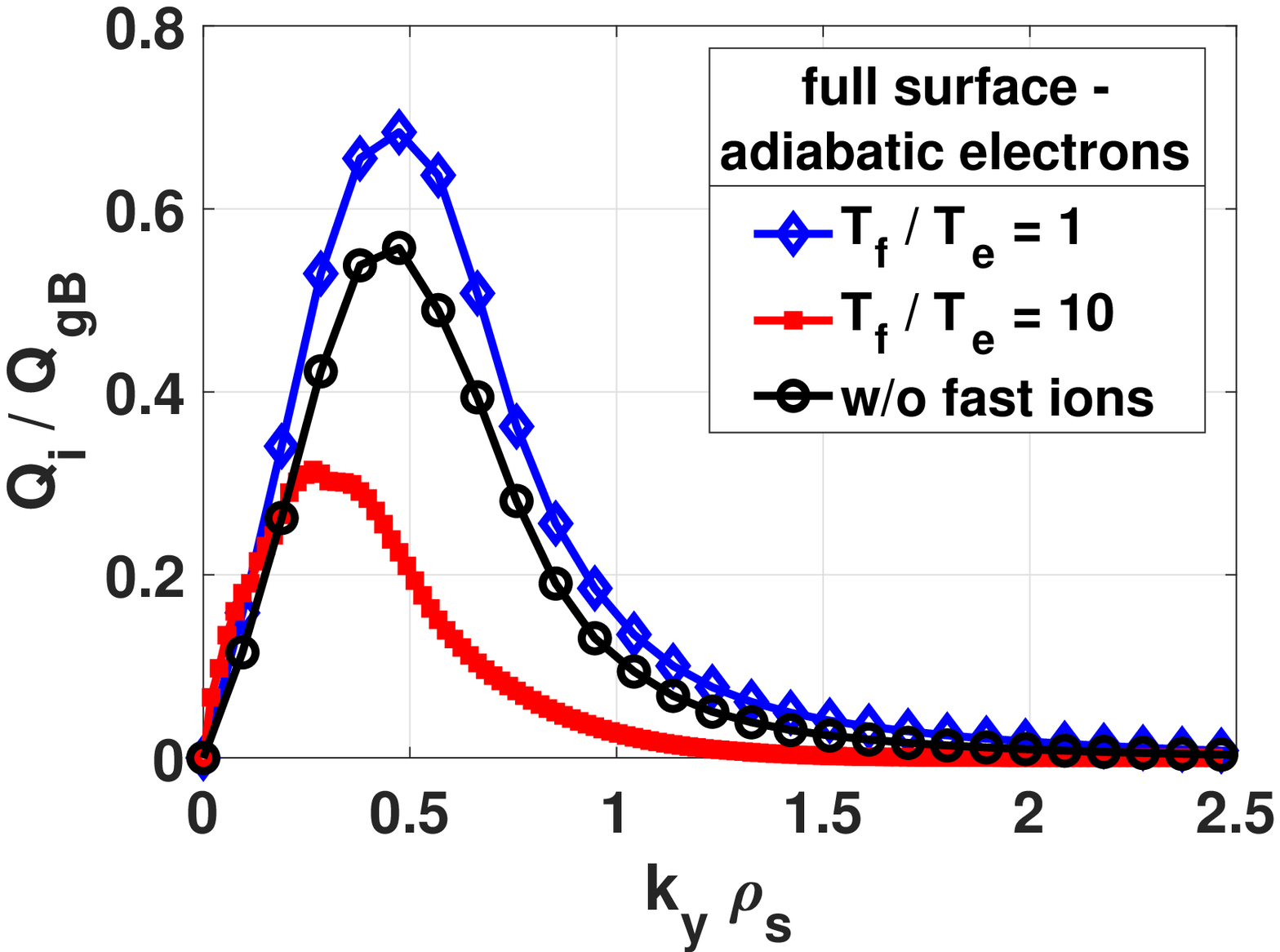}
\includegraphics[scale=0.25]{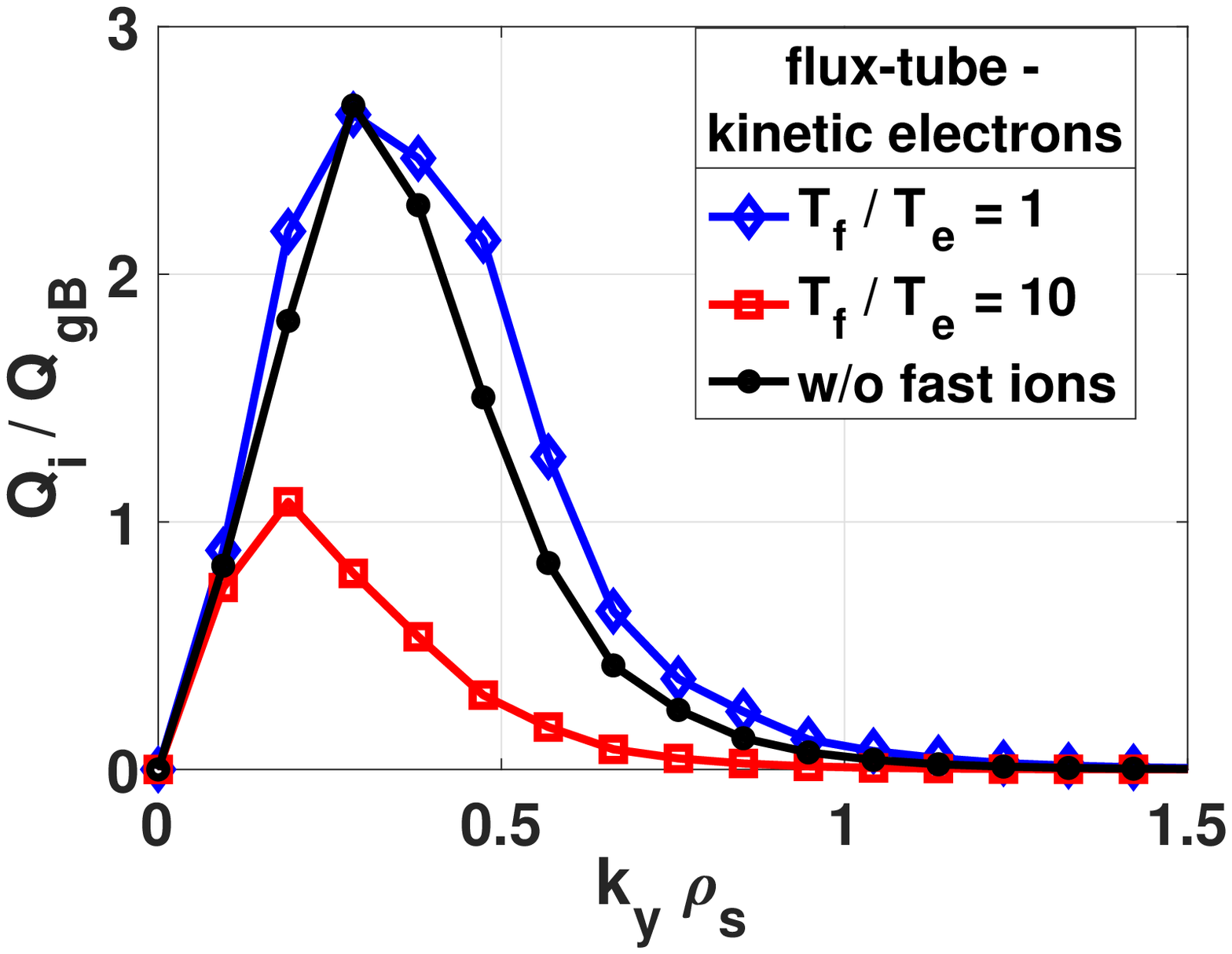}\includegraphics[scale=0.25]{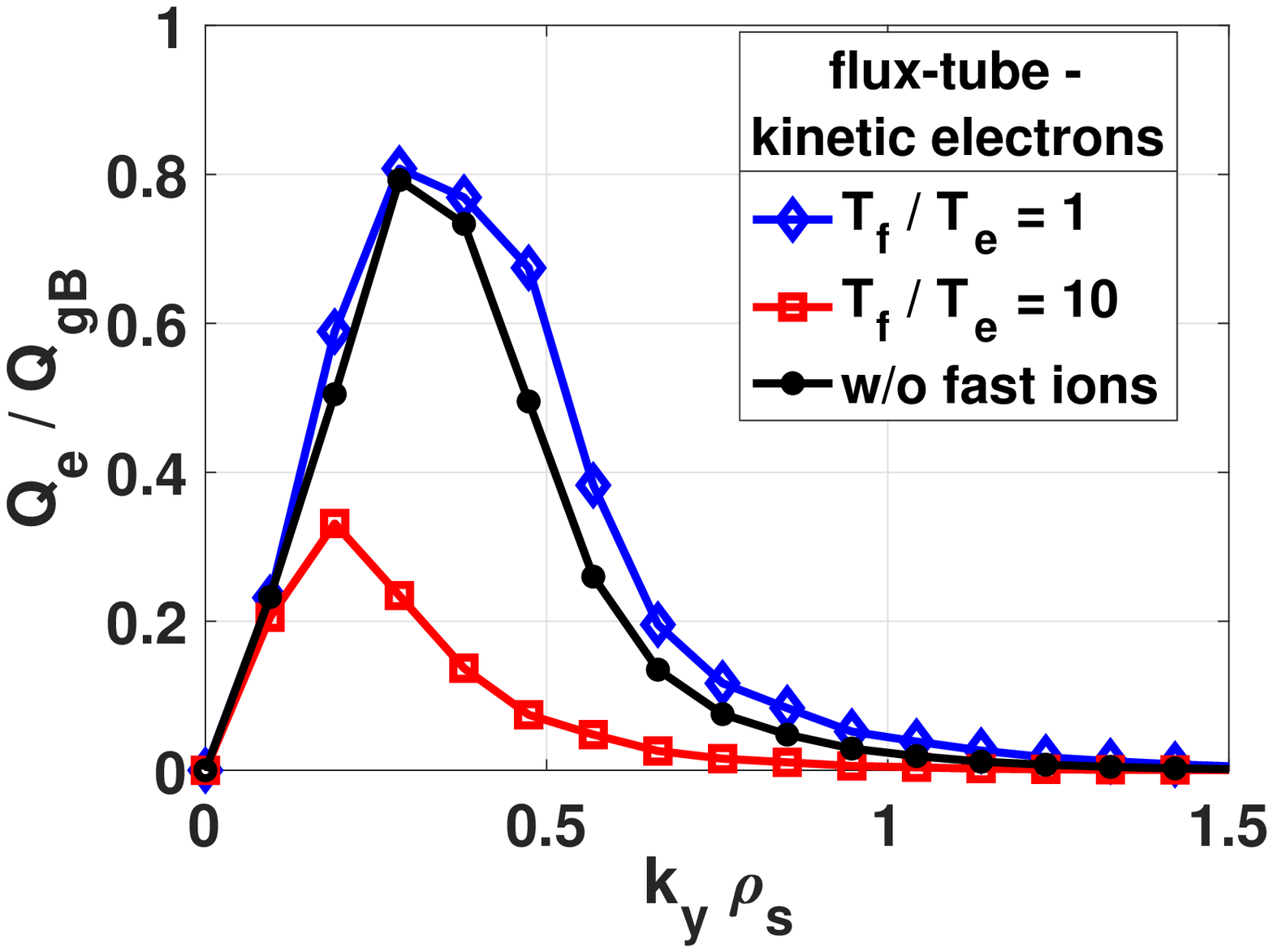}
\par\end{center}
\vspace{-0.5cm}
\caption{Main ion heat flux spectra averaged over the saturated time domain at $T_f / T_e = 1$, $T_f / T_e = 10$ and without fast ions for: the most unstable flux-tube (a) and full surface (b) simulations with adiabatic electrons; and (c) with kinetic electrons. (d) Electron heat flux spectra.} 
\label{fig:fig2}
\end{figure}

The critical role of supra-thermal ions in stellarator devices is further investigated in Fig.~\ref{fig:fig2}. 
More specifically, the $k_y \rho_s$ spectrum of the time-averaged thermal ion heat flux is studied for the adiabatic electron simulations performed for the most unstable flux-tube (Fig.~\ref{fig:fig2}a) and full surface (Fig.~\ref{fig:fig2}b) setups. Taking as reference spectra the one obtained without fast ions, a common feature arises in all of our numerical analyses.
The peak of the main ion heat flux spectra exhibits a downshift towards smaller values of $k_y \rho_s$ as the energetic particle temperature increases. It is located at $k_y \rho_s = 0.5$ for the cases without fast ions and moves, respectively, to $k_y \rho_s = 0.19$ for the most unstable flux-tube and to $k_y \rho_s = 0.28$ for the full surface simulations at $T_f = 10 T_e$. These results suggest that at a fixed temperature, each scale  undergoes a different stabilization due to fast ion effects, significantly affecting the shape of the bi-normal heat flux spectra. We generalize the previously presented results by including the full kinetic electron dynamics. In particular, the impact of the supra-thermal particles on the main ion and electron heat flux spectra is illustrated in Figs.~\ref{fig:fig2}c and ~\ref{fig:fig2}d. These findings confirm that energetic particles contribute to a substantial stabilisation of the ion scale turbulence regardless the presence of kinetic electrons. In particular, the main ion heat flux decreases by $\sim 75\%$ at $T_f/T_e = 10$. As the ITG drive is reduced by the effect of fast ions, the overall turbulent electron heat flux diminishes accordingly by $\sim 60\%$.

{\em Discussion and implications.} The findings in the present Letter show a substantial impact of energetic particles on ITG driven turbulence in W7-X. Fast ions are found to reduce the bulk energy losses as their temperature is increased, affecting the heat flux spectra differently at each bi-normal scale. Here, we demonstrate that these results are consistent with the wave-particle mechanism described above via the energy diagnostics developed in Refs.~\cite{Navarro_PRL2011,Banon_Navarro_PoP2011}. A primary focus to assess the feasibility of such a resonant interaction is the study of the phase-space structure of the energy exchanged between the plasma microinstabilities and the fast ions ($\gamma_f$).
\begin{figure}
\begin{center}
\includegraphics[scale=0.53]{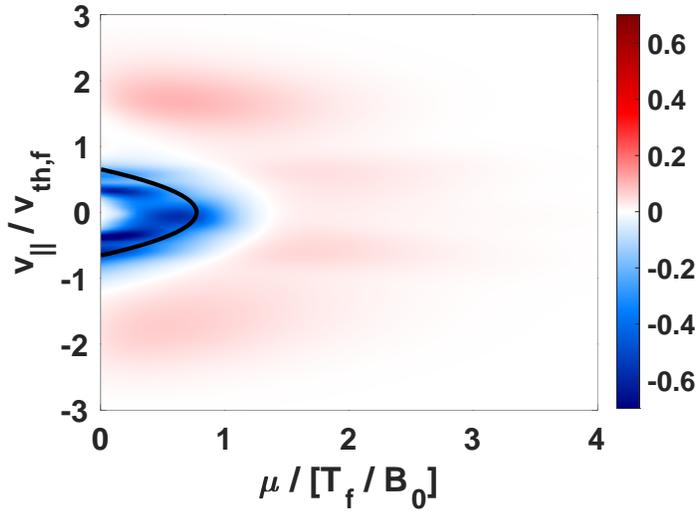}
\par\end{center}
\caption{Velocity space structure of the converged fast particle $\gamma_f$ obtained from the most unstable flux-tube simulation 
with adiabatic electrons at the bi-normal scale $k_y \rho_s = 0.5$. The energetic particle temperature is $T_f / T_e = 10$. The black contour line indicates the phase-space resonance position according to Eq.~(\ref{eq:gamma_f}). }
\label{fig:fig4}
\end{figure}
The velocity space structure $(v_\shortparallel,\mu)$ of $\gamma_f$ - averaged over the field-aligned coordinate z - is illustrated in Fig.~\ref{fig:fig4} for $T_f/T_e = 10$ at $k_y \rho_s = 0.5$, i.e. corresponding to the strongest fast ion stabilizing effect observed in Fig.~\ref{fig:fig1}. The black line of Fig.~\ref{fig:fig4} indicates the $(v_\shortparallel,\mu)$ values that satisfy the resonance condition expressed in Eq.~(\ref{eq:gamma_f}). We note the predominantly negative velocity phase-space structure of $\gamma_f$, revealing an energy redistribution from the thermal ion driven ITG microinstability to the energetic particle species. Therefore, fast ions act as an effective sink of energy at $T_f / T_e = 10$, reducing the main ITG drive and - in the more complex nonlinear case -  the overall ion-driven turbulent fluxes. It appears clear from Fig.~\ref{fig:fig4} that the strongest beneficial energetic particle contribution to $\gamma_f$ occurs in correspondence of the velocity space coordinates identified by Eq.~(\ref{eq:gamma_f}). This resonance interaction strongly affects the shape of $\gamma_f$. More precisely, an effective interaction requires that the resonance condition is satisfied in the negative region, which in Fig.~\ref{fig:fig4} occurs exactly at $E_f < 3/2$ (or equivalently at $v_\shortparallel^2+\mu B_0 < 3/2$). Only in this case, the wave-particle interaction amplifies a negative $\gamma_f$ contribution, leading to a reduction of the ITG drive and hence to a turbulence stabilisation. It is worth noting that the negative sign of $\gamma_f$ reflects inward fast particle energy fluxes, consistently with what observed in Fig.~\ref{fig:fig1}b.

\begin{figure}
\begin{center}
\includegraphics[scale=0.53]{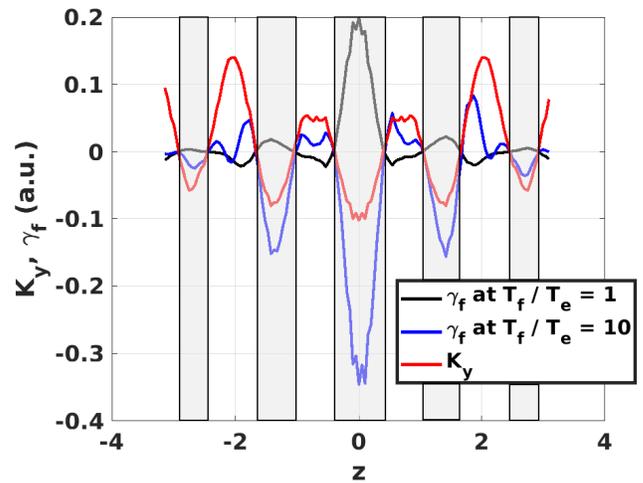}
\par\end{center}
\caption{Comparison of the field-aligned structure of the bi-normal curvature term $\mathcal{K}_y$ and the velocity space averaged $\gamma_f$ obtained from the most unstable flux-tube simulation with adiabatic electrons at $k_y \rho_s = 0.5$ and different energetic particle temperature. The area within the vertical gray boxes denotes the field-aligned values where $\mathcal{K}_y<0$.} 
\label{fig:fig5}
\end{figure}

To further corroborate these findings, the field-aligned (z) dependence of $\gamma_f$ at $T_f / T_e = 1$ and $T_f / T_e = 10$ is compared in Fig.~\ref{fig:fig5} to the bi-normal curvature term $\mathcal{K}_y$ at
$k_y \rho_s = 0.5$. 
By looking at these results, we observe a clear relation between $\gamma_f$ and $\mathcal{K}_y$. More precisely, Fig.~\ref{fig:fig5} reveals that, as the resonance interaction becomes more and more effective (i.e. by increasing the fast particles temperature), $\gamma_f$ exhibits negative values (i.e. stabilizes the ITG modes) only where $\mathcal{K}_y<0$. This feature disappears as the fast ion temperature is reduced. In contrast, for the case $T_f / T_e = 1$ shown in Fig.~\ref{fig:fig5}, the energetic particle magnetic drift $\omega_{d,f}$ is negligible compared to the linear ITG $\omega_k$ frequency and no effective resonant energy exchange is allowed. Therefore, for $T_f / T_e = 1$, the fast particle minority contributes to the development of the thermal ion driven ITG instability by adding a positive (destabilizing) contribution, which is mainly localized in the bad-curvature region at $z = 0$. 

The key role played by $\mathcal{K}_y$ in regulating the interaction between fast particles and turbulence makes this energetic particle mechanism potentially highly attractive for stellarator geometries. In such devices, $\mathcal{K}_y$ exhibits peculiar field-aligned dependencies (not present in tokamaks), which might be designed to enhance wave-particle turbulence suppression. Another possible optimization parameter found in our analyses is the fast particle temperature. The optimal value is found at $T_f / T_e = 10$. While this seem to be a localized sweet spot in flux-tube simulations, the more realistic full surface results show a broader range of temperatures where this stabilisation is significant, making this condition more accessible experimentally.

{\em Conclusions.} In the present Letter, we show that supra-thermal ions - e.g., from ICRF heating - can strongly suppress turbulent transport in optimized stellarators. This is demonstrated via a combination of analytic theory and state-of-the-art gyrokinetic turbulence simulations for W7-X. We find that the key mechanism for this suppression is a resonant wave-particle interaction between fast ions and ITG modes. This is the first time that such a mechanism is being proposed for stellarators. These results have important implications for the further optimization of such devices, where this resonant mechanism may be exploited systematically to access new improved confinement scenarios with steep temperature profiles, as a result of the turbulence stabilization. Obviously, these findings strongly motivate the use of ICRF heating systems in present-day optimized stellarator devices, such as W7-X, as a means of creating high performance discharges.

The simulations presented in this work were performed at the Cobra HPC system at the Max Planck Computing and Data Facility (MPCDF), Germany. Furthermore, we acknowledge the CINECA award under the ISCRA initiative, for the availability of high performance computing resources and support. The authors would like to thank T. G{\"o}rler,  J. P. Martin Collar, and G. G. Plunk for all the stimulating discussions, useful suggestions and comments.

\end{document}